\documentclass[12pt]{article}
\usepackage{epsf}
\usepackage{amsmath}
\usepackage{graphics}

\renewcommand{\thefootnote}{\#\arabic{footnote}}

\newcommand{\lesssim}{ \mathop{}_{\textstyle \sim}^{\textstyle <} }

\begin{document}

\setcounter{footnote}{0}
\begin{titlepage}

\begin{center}

\hfill astro-ph/0407631\\
\hfill July 2004\\

\vskip .5in

{\Large \bf
Making waves on CMB power spectrum \\
and  inflaton dynamics
}

\vskip .45in

{\large
Masahiro Kawasaki, Fuminobu Takahashi and Tomo Takahashi
}

\vskip .45in

{\em
Institute for Cosmic Ray Research, \\
University of Tokyo, Kashiwa 277-8582, Japan
}

\end{center}

\vskip .4in

\begin{abstract}
We discuss cosmic microwave background anisotropies in models with an
unconventional primordial power spectrum. In particular, we consider
an initial power spectrum with some ``spiky'' corrections.
Interestingly, such a primordial power spectrum generates ``wavy''
structure in the CMB angular power spectrum.  
\end{abstract}
\end{titlepage}

\renewcommand{\thepage}{\arabic{page}}
\setcounter{page}{1}
\renewcommand{\thefootnote}{\#\arabic{footnote}}

Result of the first year Wilkinson microwave anisotropy probe (WMAP)
\cite{Bennett:2003bz} is almost consistent with the following
assumption \cite{Spergel:2003cb}:the present universe consists of 5 \%
baryon, 23 \% cold dark matter (CDM) and 72 \% dark energy. The
primordial power spectrum is almost scale invariant with adiabatic
fluctuations.  These assumptions form so-called the standard scenario
of cosmology.

Although the fit to the WMAP data with the standard scenario seems to
be excellent, there may be some deviations from the standard scenario
in the angular power spectrum given by WMAP.  One of them is that the
angular power spectrum seems to be suppressed at low multipoles
compared to the prediction of the standard cosmological scenario~\footnote{
In fact, this suppression was already observed by COBE and 
this issue was discussed in Refs.~\cite{cobe_lowCl}. }.

Since the cosmic variance error is very large at low multipole region,
we may say that the suppression is just a statistical
fluctuation. However, if we take this seriously, it may suggest some
new physics.  There are many works proposing mechanisms to
explain this suppression \cite{suppressed_low}.

The other one is that, if you look at the WMAP angular power spectrum
carefully, there is ``wavy'' structure around multipole region from $
l\sim \mathcal{O}(10)$ up to the first peak.  In particular, there
seems to be a dip around the first peak.  Although this feature is not
robust at this point, if it is an intrinsic one, it may suggest some
unusual physics since this kind of structure cannot be explained by
the conventional scenario.  There have been some works inspired by
this feature.  
In Ref. \cite{Kogo:2003yb}, using a cosmic inversion
method, reconstruction of the primordial power spectrum which can 
accommodate this feature is discussed.
The authors of Refs.  \cite{Martin:2003sg,Martin:2004iv,Martin:2004yi}
considered a primordial power spectrum with oscillatory corrections
which can be generated from trans-Planckian physics. They showed that,
with such an initial power spectrum, we can improve the fit to the
WAMP data compared with the best-fit value of the standard scenario.

Although we may not be at the position where we should take this
result seriously, it is interesting to consider some models which give
the wavy structure on the CMB angular power spectrum.  In this letter,
we discuss a mechanism which generates such structure in the CMB power
spectrum.  Here we consider a scenario where the primordial
fluctuation is modified from the standard form.
If we assume an initial power spectrum with some ``spiky''
parts, wavy structure of the CMB power spectrum can appear.  
The initial power spectrum we assume is written as 
\begin{equation}
P_s = A_s k^{n-1} \left( 
1 + \sum_i \alpha_i \exp \left[ -\frac{(k-k_{ci})^2}{2 \sigma_{i}^2} \right] 
\right),
\label{eq:P_s}
\end{equation}
where $A_s$ and $n$ are the amplitude of primordial fluctuation and
the spectral index as usually assumed.  In the spectrum, ``spiky''
corrections are added to the standard power spectrum which can be
parameterized by $\alpha_i, k_{ci}$ and $\sigma_{i}$ for the $i$-th
spike.  We assume $\alpha_i \ge -1$ to keep the initial power spectrum
$P_s$ positive.  $k_{ci}$ and $\sigma_i$ represent the position of
spike and its width respectively.  These parameters can be determined
by models of the mechanism generating the primordial fluctuation.  Here
we consider a situation where the width of the Gaussian function is
very narrow.  This kind of primordial spectrum may arise from a
particular kind of inflaton dynamics, which will be discussed later.

\begin{figure}
    \centerline{\scalebox{1.2}{\includegraphics{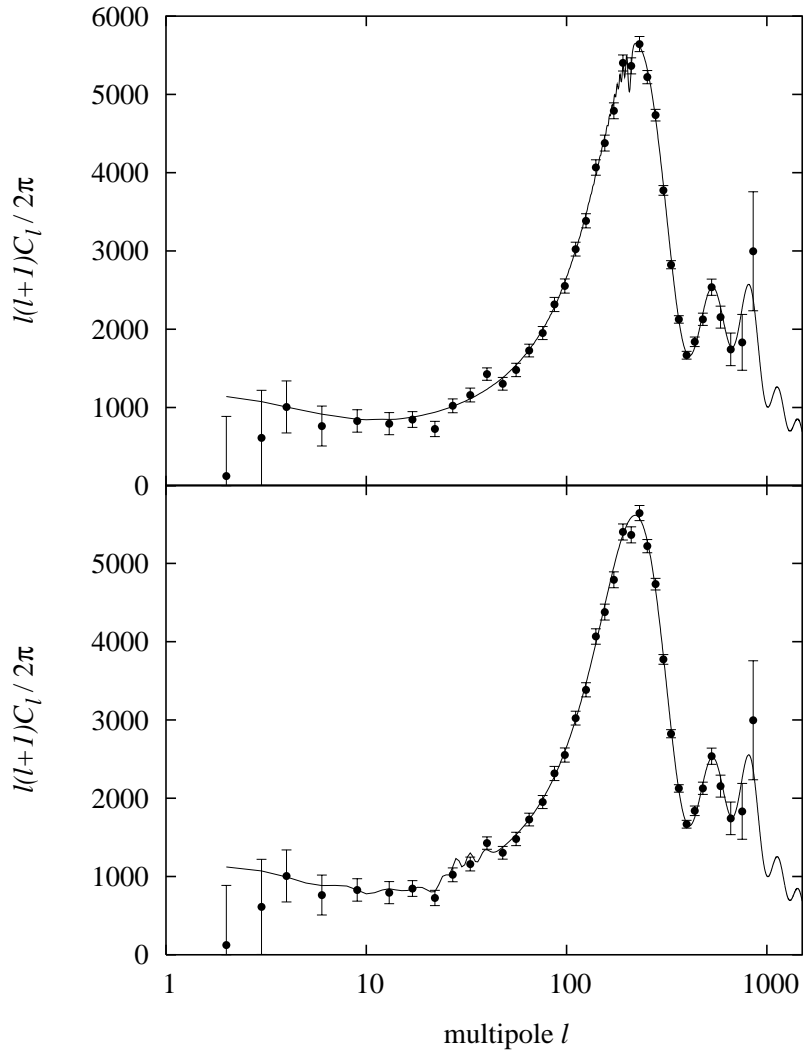}}}
	\caption{The CMB angular power spectrum with the primordial
	spectrum Eq.\ (\ref{eq:P_s}).  The model parameters and
	cosmological parameters are given in the text. The binned WMAP 
	data is also plotted.}
\label{fig:cl}
\end{figure}

First we discuss effects of the spiky corrections in the initial power
spectrum.  In Fig.\ \ref{fig:cl}, we show the resultant CMB angular
power spectrum from the initial power spectrum given by Eq.\
(\ref{eq:P_s}).  In the upper panel of the figure, we assume a single
``dip'' which is parameterized as $\alpha = -0.9$, $k_{c} = 2.2 \times
10^{-2} h ~{\rm Mpc}^{-1}$, and $\sigma=8.7 \times 10^{-5} h~{\rm
Mpc}^{-1}$. In the lower panel of the figure, we assumed a single
spike and a single dip  in the initial power spectrum with the
parameters $\alpha_1 = 3.0, k_{c1} = 4.3 \times 10^{-3} h ~{\rm
Mpc}^{-1}, \sigma_{1}= 7.6 \times 10^{-5} h~{\rm Mpc}^{-1}$ and
$\alpha_2 = -0.9, k_{c2} = 2.5 \times 10^{-3} h ~{\rm Mpc}^{-1},
\sigma_{2}= 1.4 \times 10^{-4} h~{\rm Mpc}^{-1}$.  The cosmological
parameters are chosen to be $\Omega_m h^2=0.145, \Omega_b h^2= 0.023,
h = 0.69, n_s = 0.97$ and $\tau=0.116$ which give the minimum value of
$\chi^2$ in the power-law $\Lambda$CDM model~\footnote{
Although these cosmological parameters are a bit different from those 
given by WMAP \cite{Spergel:2003cb}, this choice of cosmological parameters
gives a better fit to the data with $\chi^2 = 1428.6$ \cite{Ichikawa}.}.
Here a flat universe and no tensor mode are assumed.  Very
interestingly, with the parameter in the upper panel, you can see a
dip around the first peak, while 
wavy structure around $l\sim 50$ appears in the lower panel. 
Accordingly, the fit to the
WMAP data becomes better. We calculated $\chi^2$ using the code
provided by WMAP \cite{Verde:2003ey} and obtained smaller $\chi^2$ by
$\Delta \chi^2 \sim -10$ compared to the value for the best-fit
parameter of the power-law $\Lambda$CDM model in both panels in 
Fig.~\ref{fig:cl}~\footnote{
For the case with the upper panel of the figure, assuming a dip
introduces three additional parameters thus the degrees of freedom
increases by 3.  Similarly, in the case with the lower panel of the
figure, the degrees of freedom increases by 6 since we assume a dip
and a spike.}
.

\begin{figure}
    \centerline{\includegraphics{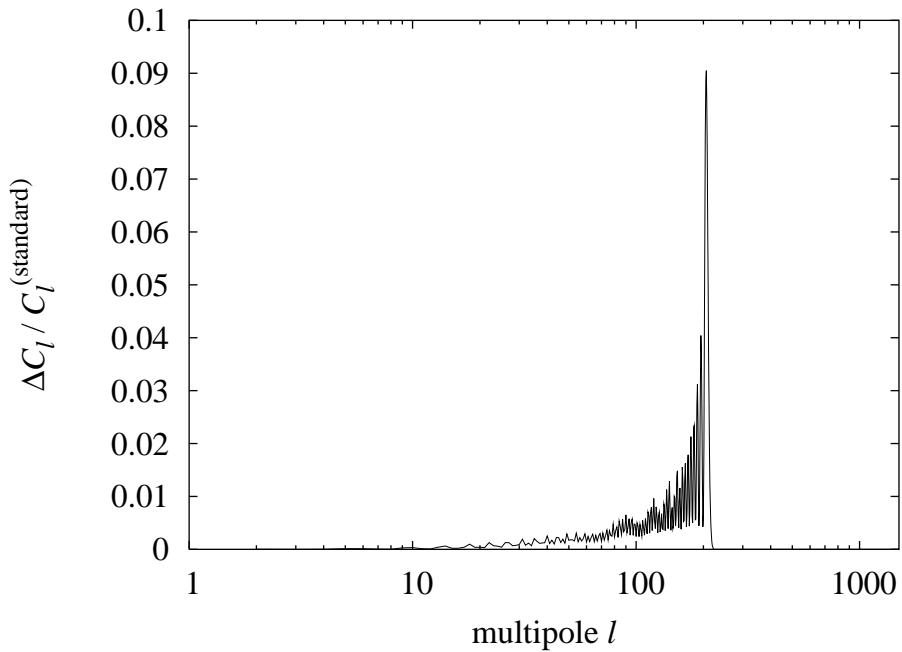}} 
	\caption{The difference in $C_l$ between with and without
	spiky correction terms. The model and cosmological parameters
	are the same as in the upper panel of Fig.\ \ref{fig:cl}.}
\label{fig:delta_cl}
\end{figure}

The reason why such structure arises is simple.  The angular power
spectrum can be written as 
\begin{equation}
C_l = \frac{2}{\pi} \int_0^\infty dk~ k^2 P_s(k) |\Theta_l(k,\eta_0)|^2
\end{equation}
where $\Theta_l(k,\eta)$ is the transfer function and $P_s$ is the
primordial power spectrum.  $\eta$ is the conformal time and the
subscript $0$ represents a quantity defined at the present time.  The
transfer function is given by \cite{Dodelson}
\begin{eqnarray}
\Theta_l (k,\eta_0) &=& 
\int_0^{\eta_0} d\eta~ g(\eta)  (\Theta_0 (k,\eta)+  \Psi(k,\eta) ) j_l(k(\eta_0 - \eta)) 
\notag \\
&&- \int_0^{\eta_0} d\eta~ g(\eta)  \frac{i v_b (k, \eta)}{k} 
\frac{d}{d\eta} j_l(k(\eta_0 - \eta)) 
\notag \\
&& +  \int_0^{\eta_0} d\eta~ e^{-\tau} 
\left[ \dot{\Phi}(k,\eta) - \dot{\Psi}(k,\eta) \right] j_l(k(\eta_0 - \eta)) 
\label{eq:transfer}
\end{eqnarray}
where $j_l$ is a spherical Bessel function of degree $l$, and
$v_b$ is the velocity of baryon. $\Psi$ and $\Phi$
are the gravitational potential and the curvature perturbation,
respectively.  $g(\eta)$
is the visibility function which has a peak around $\eta_*$ where $*$
represents the epoch of the recombination.  Since we are considering
the case where $\sigma_{i}$ is very small (i.e., the width is very
narrow), we can replace the second term of Eq.\ (\ref{eq:P_s}) with
delta functions $\sum_i \delta (k-k_{ci})$.  Approximating with the
delta function in the initial power spectrum Eq.\ (\ref{eq:P_s}), 
we can write $C_l$ as
\begin{eqnarray}
C_l \simeq C_l^{\rm (standard)} 
&+& \sum_i \alpha_i
\left|F^{(1)}_l(k_{ci}, \eta_0, \eta_*) j_l(k_{ci} (\eta_* - \eta_0))
%\right. \nonumber \\ &&\left.
+F^{(2)}_l(k_{ci}, \eta_0, \eta_*) \frac{d}{d\eta}j_l(k_{ci} (\eta_* - \eta_0))
\right|^2, \nonumber \\
\label{eq:Cl_2}
\end{eqnarray}
where $F^{(1,2)}_l(k_{ci}, \eta_0, \eta_*)$ are functions which vary slowly
with $l$ and $C_l^{\rm (standard)}$ is a CMB power spectrum calculated
by using the standard initial power spectrum.  As you can see from
Eq.\ (\ref{eq:Cl_2}), when the spiky correction term in the initial
power spectrum can give sizable contribution to $C_l$, the form of
$C_l$ exhibits the oscillatory structure of $j_l$. This is the reason
why the wavy structure of $C_l$ can be realized assuming the initial
power spectrum given by Eq.\ (\ref{eq:P_s}).  To see the effects of a
spherical Bessel function, we plot $\Delta C_l/C_l^{\rm (standard)} = (C_l^{\rm
(standard)} - C_l)/C_l^{\rm (standard)}$ in Fig.\ \ref{fig:delta_cl} with
the same model and cosmological parameters in the upper panel of Fig.\
\ref{fig:cl}.  When $\alpha_i$ is negative, the second term in Eq.\
(\ref{eq:Cl_2}) gives a negative contribution to $C_l$, thus, in this
case, a dip can appear as in the upper panel of Fig.\ \ref{fig:cl}.
For a positive $\alpha_i$ case, the second term is added to give
oscillatory power spectrum.  Without a spiky part in the initial power
spectrum as it is usually assumed, oscillatory structure of the
spherical Bessel function is smeared out by contributions from
superposition of $k-$modes around the corresponding multipole
region. Hence we do not usually see such structure. Notice also that
if the width of the Gaussian part in the initial power spectrum of
Eq.\ (\ref{eq:P_s}) is not narrow, the oscillatory structure of a
spherical Bessel function is also smeared out, thus we cannot see the
wavy structure on the CMB power spectrum.
Then how narrow should the width of the spiky structure be to give a
distinctive feature to $C_l$? Noting that the spherical Bessel function $j_l(z)$
has the highest peak with the width $\Delta z \sim l^{1/3}$ around $z_* \sim l$,
the following inequality needs to be satisfied in order to make wavy structure
around multipole $l$:
\begin{equation}
\frac{\sigma_i }{ k_{ci}} \lesssim \frac{\Delta z}{z_* }\sim l^{-2/3}.
\end{equation}
This inequality is satisfied for the model parameters we have adopted in Fig.~\ref{fig:cl}.

Now we discuss possible models which generate the initial power
spectrum of the form of Eq.\ (\ref{eq:P_s}).  Usually, the primordial
density fluctuation is represented by the amplitude of the density
perturbation at the horizon crossing $\delta_H(k)$ which can be
approximately written in the spatially flat gauge as
\cite{KolbTurner},
\begin{equation}
\delta_H (k) \sim \frac{H^2}{5\pi \dot{\phi}},
\label{eq:deltaH}
\end{equation}
where $H$ is the Hubble parameter during inflation and $\phi$ is an
inflaton field.  The dot represents derivatives with respect to time.
Using $\delta_H$, the initial power spectrum can be written as $P_s
(k)\equiv \delta_H^2(k)$.  From Eq.\ (\ref{eq:deltaH}), we can see
that, if the inflaton field almost stops instantaneously, $\delta_H$
gets enhanced at corresponding scale $k_c$, which may make some spiky
structure in the power spectrum~\footnote{
Note that the power spectrum does not diverge even in the limit of $\dot{\phi}
\rightarrow 0$~\cite{Seto:1999jc}.
}.  Inversely, if the inflaton field moves very fast instantaneously,
$\delta_H$ gets suppressed at corresponding scale $k_c$. In this case,
this makes some dips in $\delta_H(k)$. Thus the primordial power
spectrum in the form of Eq.\ (\ref{eq:P_s}) may be generated from an
unusual inflaton dynamics.  However, this simple argument is not
enough to evaluate the primordial power spectrum.
For more detailed investigation of this issue, see Ref. \cite{Elgaroy:2003hp}.

In summary, we considered the primordial power spectrum with some
spikes and/or spiky dips. Assuming this kind of  initial power
spectra, wavy structure appears in the CMB angular power spectrum,
which may be inferred from the first year WMAP observation.  Such wavy
structure may originate to the oscillatory structure of a spherical
Bessel function. When there is a spike in the initial power spectrum
and its width is very narrow, for a fixed $l$, only the small number
of $k$-modes is picked up to give contributions to $C_l$ with
oscillatory structure of the spherical Bessel function. If this mode
can give sizable contribution to $C_l$, the oscillatory behavior of
the spherical Bessel function directly affects the form of $C_l$. When
the width of the spike is not so narrow, several modes can contribute
to $C_l$. Thus oscillatory structure is smeared out by the projection
effect as the conventional case.

\bigskip
\bigskip

\noindent
{\bf Acknowledgment:} We would like to thank K.~Ichikawa for useful
discussion.  F.T. and T.T would like to thank the Japan Society for
Promotion of Science for financial support.  This work is partially
supported by the JSPS Grand-in-Aid for Scientific Research
No. 14540245 (M.K.).  We acknowledge the use of CMBFAST
\cite{Seljak:1996is} for numerical calculations.

\end{document}